\RequirePackage{ifpdf}
\ifpdf 
\documentclass[pdftex]{sigma}
\else
\documentclass{sigma}
\fi

\begin{document}
\allowdisplaybreaks

\renewcommand{\PaperNumber}{028}

\FirstPageHeading

\ShortArticleName{Application of the Gel'fand Matrix Method to the
Missing Label Problem}

\ArticleName{Application of the Gel'fand Matrix Method\\ to the
Missing Label Problem \\ in Classical Kinematical Lie Algebras}

\Author{Rutwig CAMPOAMOR-STURSBERG} 
\AuthorNameForHeading{R. Campoamor-Stursberg}

\Address{Departamento Geometr\'{\i}a y Topolog\'{\i}a,
Fac. CC. Matem\'aticas U.C.M.,\\
Plaza de Ciencias 3, E-28040 Madrid, Spain}

\Email{\href{mailto:rutwig@mat.ucm.es}{rutwig@mat.ucm.es}}

\ArticleDates{Received November 06, 2005, in f\/inal form February
14, 2006; Published online February 28, 2006}

\Abstract{We brief\/ly review a matrix based method to compute the
Casimir operators of Lie algebras, mainly certain type of
contractions of simple Lie algebras. The versatility of the method
is illustrated by constructing matrices whose characteristic
polynomials provide the invariants of the kinematical algebras in
(3+1)-dimensions. Moreover it is shown, also for kinematical
algebras, how some reductions on these matrices are useful for
determining the missing operators in the missing label problem
(MLP).}

\Keywords{Casimir operator; characteristic polynomial; 
Lie algebra; missing label; kinematical group}

\Classification{17B05; 81R05}

\section{Introduction}

Casimir operators of Lie algebras were originally introduced in
the frame of representation theory of semisimple algebras, and
were soon recognized as a powerful tool for the structural
analysis. They were determined by means of the universal
enveloping algebra of a Lie algebra, and successive applications
of Lie groups and algebras to other mathematical and physical
problems led to alternative methods to compute these invariants,
like the orbit method or the formulation in terms of
dif\/ferential equations, which allow to generalize the concept of
invariant functions on Lie algebras beyond the pure algebraic
frame.

In 1950 I.M.~Gel'fand \cite{Ge} showed how to use the generic
matrix of standard representation of orthogonal groups to
determine the Casimir operators of the algebra by means of
characteristic polynomials. More specif\/ically, for the
pseudo-orthogonal Lie algebra $\frak{so}(p,q)$ $(N=p+q)$ given by
the $\frac{1}{2}N(N-1)$ operators $E_{\mu\nu}=-E_{\nu\mu}$ and
having brackets:
\begin{gather*}
\left[  E_{\mu\nu},E_{\lambda\sigma}\right]
=g_{\mu\lambda}E_{\nu\sigma
}+g_{\mu\sigma}E_{\lambda\nu}-g_{\nu\lambda}E_{\mu\sigma}-g_{\nu\sigma
}E_{\lambda\mu},
\end{gather*}
where $g={\rm diag}\left(  1,\ldots,1,-1,\ldots,-1\right)$, the
Casimir operators were obtained with the formula~\cite{Ge}:
\begin{gather*}
P(\lambda)=\left|  M_{p,q}-\lambda\,\mathrm{Id}_{N}\right|  =\lambda^{N}%
+\sum_{k=1}^{N}C_{k}\lambda^{N-k},
\end{gather*}
$M_{p,q}$ being the matrix
\begin{gather}
M_{p,q}=\left(
\begin{array}
[c]{ccccc}%
0 & \cdots & -g_{jj}e_{1j} & \cdots & -g_{NN}e_{1N}\\
\vdots &  & \vdots &  & \vdots\\
e_{1j}  &\cdots & 0 & \cdots & -g_{NN}e_{jN}\\
\vdots &  & \vdots &  & \vdots\\
e_{1N}  & \cdots & g_{jj}e_{jN}  & \cdots & 0
\end{array}
\right)  \label{ST}%
\end{gather}
and ${\rm Id}_{N}$ the identity matrix of order $N$. Once
symmetrized, the functions $C_{k}$ provide the classical operators
in the enveloping algebra of $\frak{so}(p,q)$. This direct matrix
approach seems natural, taking into account that the eigenvalues
of Casimir operators are essential for the labelling of
representations of the Lie algebra. By application of the
classical theory of semisimple Lie algebras, similar formulae can
be developed for other semisimple Lie algebras \cite{Pe}.  It is
natural to ask whether the essence of the Gel'fand method can be
generalized to other non-semisimple Lie algebras, by means of
extended matrices that possibly correspond to some representation
of the algebra. The two main interesting cases are:
\begin{enumerate}
\vspace{-2mm}\itemsep=0pt \item Generalized In\"on\"u--Wigner
contractions of (semi)simple Lie algebras. We have inhomogeneous
algebras as a special case.

\item Semidirect products of simple and Heisenberg Lie algebras.
\vspace{-2mm}
\end{enumerate}

The method usually employed to determine the invariants of a Lie
algebra consists in solving of a system of partial dif\/ferential
equations (PDEs) related to a realization of the Lie algebra as
dif\/ferential operators \cite{Fu,Wi}. If $\left\{
X_{1},\ldots,X_{n}\right\} $ is a basis of $\frak{g}$ and $\{
C_{ij}^{k}\}  $ is the structure tensor, the realization of
$\frak{g}$ in the space $C^{\infty}( \frak{g}^{\ast}) $ is given
by:
\begin{gather*}
\widehat{X}_{i}=-C_{ij}^{k}x_{k}\partial_{x_{j}},
\end{gather*}
where $\left[  X_{i},X_{j}\right]  =C_{ij}^{k}X_{k}$ $\left( 1\leq
i<j\leq n,\;1\leq k\leq n\right)  $, $\left\{
x_{1},\ldots,x_{n}\right\} $ is the corresponding dual basis,
$\frak{g}^{*}$ being the dual space to $\frak{g}$. Invariants of
the coadjoint representation are functions on the generators
$F\left( X_{1},\ldots,X_{n}\right)  $ of $\frak{g}$ such that
$\left[ X_{i},F\left( X_{1},\ldots,X_{n}\right) \right] =0$, and
are determined by solving the system of PDEs:
\begin{gather}
\widehat{X}_{i}F\left(  x_{1},\ldots ,x_{n}\right)
=-C_{ij}^{k}x_{k}\frac{\partial F}{\partial x_{j}}\left(
x_{1},\ldots ,x_{n}\right)  =0,\qquad 1\leq i\leq n. \label{sys}
\end{gather}
Classical Casimir operators are recovered with replacing the
variables $x_{i}$ by the corresponding gene\-rator $X_{i}$
(possibly after symmetrizing). The cardinal $\mathcal{N}\left(
\frak{g}\right)  $ of a maximal set of independent solutions is
described in terms of the following formula:
\begin{gather}
\mathcal{N}\left(  \frak{g}\right)
=\dim\,\frak{g}-\mathrm{rank}\,A\left(\frak{g}\right)
,\label{BB}
\end{gather}
where $A\left(\frak{g}\right)  $ is the matrix representing the
commutator table of $\frak{g}$ over a given basis, i.e.,
\begin{gather*}
A(\frak{g})=\big(  C_{ij}^{k}x_{k}\big).
\end{gather*}

In this work we brief\/ly review some recent work on matrix
procedures that generalize the Gel'fand formula for various types
of Lie algebras, specif\/ically inhomogeneous Lie algebras.
Applications to the computation of missing label operators are
also given.

\section{Contractions of simple Lie algebras}

Let $\frak{g}$ be a simple Lie algebra of rank $k$ such that the
Casimir operators of $\frak{g}$ are given by the characteristic
polynomial $P\left(  T\right)  $ of a matrix
\[
A=\sum_{i}\Gamma_{1}\left(  X_{i}\right)x_{i}  ,
\]
\newpage

\noindent
where $\Gamma_{1}$ is equivalent to the $N$-dimensional standard
representation of $\frak{g}$. Suppose further that the Lie algebra
$\frak{g}^{\prime}$ is a contraction of $\frak{g}$ and that
$\mathcal{N}(\frak{g})=\mathcal{N}(\frak{g}^{\prime})$. To compute
the Casimir operators of $\frak{g}^{\prime}$ using a matrix
extension of $A$, the following procedure by steps has been
proposed~\cite{C46}:

\begin{enumerate}\vspace{-2mm}\itemsep=0pt
\item Take the matrix $A=\sum_{i}\Gamma_{1}\left(
X_{i}\right)x_{i}$.

\item Let $\Psi_{\epsilon}$ be the automorphism giving the
contraction\footnote{By this we mean that the limit
$\left[X,Y\right]_{\infty}:=\lim\limits_{\epsilon\rightarrow
\infty}\Psi_{\epsilon}^{-1}\left[\Psi_{\epsilon}(X),\Psi_{\epsilon}(Y)\right]$
exists for any $X,Y\in\frak{g}$. This equation def\/ines a Lie
algebra $\frak{g}^{\prime}$ called the contraction of $\frak{g}$
(by $\Psi_{\epsilon}$). } and  consider the matrix
$A^{\prime}=\sum_{i}\Gamma_{1}\left(X_{i}^{\prime})\right)x_{i}^{\prime}$
giving the invariants of $\frak{g}$ over the transformed basis
$\left\{X_{i}^{\prime}:=\Psi_{\epsilon}(X_{i})\right\}$.

\item Let $P_{\epsilon}(T):=\left| A^{\prime}-T\,{\rm
Id}_{N}\right|$ and $\alpha:=\deg_{\epsilon}P_{\epsilon}(T)$. Take
the limit
$P(T):=\lim\limits_{\epsilon\rightarrow\infty}\frac{1}{\epsilon^{\alpha}}P_{\epsilon}(T)$.

\item Rewrite $P(T)$ as a linear combination of determinants.

\item  Def\/ine $I_{j}=C_{j}$ as the polynomial coef\/f\/icients
of $P\left( T\right)  $.

\item  Check the functional independence of the functions $I_{j}$.

\item  If necessary, symmetrize the functions $I_{j}$ in order to
recover the Casimir operators of $\frak{g}^{\prime}$.\vspace{-2mm}
\end{enumerate}

The  validity of the method is formally proved using the
contraction explicitly. If $\Psi_{\varepsilon}$ is the
automorphism of $\frak{g}$ def\/ining the contraction, then the
matrix
\[
A_{k}^{\prime}=\sum_{i}\Gamma_{k}\left(X_{i}^{\prime}\right)x_{i}^{\prime}
\]
expresses the Casimir operators of $\frak{g}$ over the transformed
basis. Developing the corresponding characteristic polynomial
$P(T)$ and taking into account the limit,  after some algebraic
manipulation we can obtain an expression of $P(T)$ that does not
involve the contraction parameter $\epsilon$ anymore. By
transitivity of contractions, the same algorithm can be applied
formally to the contraction of reductive algebras.

\medskip

For Lie algebras of the type
$w\frak{s}=\frak{s}\overrightarrow{\oplus}_{\Gamma}\frak{h}_{N}$,
i.e., semidirect products of a simple and a Heisenberg Lie
algebra, the argument remains valid for the contraction
$w\frak{s}\rightsquigarrow
\frak{s}\overrightarrow{\oplus}_{\Gamma}(2N+1)L_{1}$. The
structure of representations of simple Lie algebras compatible
with Heisenberg algebras was analyzed in \cite{C45}. Depending on
this structure, we obtain by contraction either an inhomogeneous
Lie algebra or a general af\/f\/ine algebra. The f\/irst step is
to compute invariants of $w\frak{s}$. This is done by constructing
a copy of the Levi part $\frak{s}$ in the enveloping algebra of
$w\frak{s}$\footnote{These variables can be found analyzing the
noncentral Casimir operator of the subalgebras
$\frak{s}^{\prime}\overrightarrow{\oplus}_{\Gamma}\frak{h}_(N)$,
where $\frak{s}^{\prime}$ is a simple subalgebra of rank one of
$\frak{s}$ \cite{Que,C42}.}, which can be used as the simple
algebra to which the preceding procedure is applied.

\begin{remark}
In general, the shape of $P(T)$ will depend essentially on the
structure of the standard representation and the contraction
considered, and possibly involves matrices depending on $T$. In
particular, it will be dif\/ferent for the cases where this
 representation is by real or complex matrices, i.e., if it is of f\/irst or second genus~\cite{Iw}.
This fact can also originate some dependence problems.
\end{remark}

\subsection{Inhomogeneous symplectic algebras}

As an example of a general class to which the algorithm applies,
we consider the inhomogeneous symplectic Lie algebras
$I\frak{sp}(2N,\mathbb{R})$ and the matrix formula obtained for
them in \cite{C46}. Over the basis $\left\{
X_{i,j},X_{-i,j},X_{i,j},P_{i},Q_{i}\right\}$ ($1\leq i,j\leq N)$
the brackets of $I\frak{sp}\left(  2N,\mathbb{R}\right)  $ are
given by:
\begin{gather*}
\left[  X_{i,j},X_{k,l}\right]  =\delta_{jk}X_{il}%
-\delta_{il}X_{kj}+\varepsilon_{i}\varepsilon_{j}\delta_{j,-l}X_{k,-i}
-\varepsilon_{i}\varepsilon_{j}\delta_{i,-k}X_{-j,l},\\
\left[  X_{i,j},P_{k}\right]  =\delta_{jk}P_{i},\qquad \left[
X_{i,j}
,Q_{k}\right]  =-\delta_{ik}Q_{j},\\
\left[  X_{-i,j},Q_{k}\right]
=-\delta_{jk}P_{i}-\delta_{ik}P_{j},\qquad \left[
X_{i,-j},P_{k}\right]  =\delta_{ik}Q_{j}+\delta_{kj}Q_{i}.
\end{gather*}

\begin{proposition}
Let $N\geq2$. Then the Casimir operators $C_{2k}$ of
$I\frak{sp}\left(
2N,\mathbb{R}\right)  $ are given by the coefficients of the polynomial%
\begin{gather*}
\left|  C-T\,\mathrm{Id}_{2N+1}\right|  +\left|
C_{2N+1,2N+1}-T\,\mathrm{Id}_{2N}\right|
T=\sum_{k=1}^{N}C_{2k+1}T^{2N+1-2k},
\end{gather*}
where
\begin{gather*}
C=\left(
\begin{array}
[c]{ccccccc}%
x_{1,1} & \cdots  & x_{1,N} & -x_{-1,1} & \cdots  & -x_{-1,N} & p_{1}T\\
\vdots &  & \vdots & \vdots &  & \vdots & \vdots\\
x_{N,1} & \cdots  & x_{N,N} & -x_{-1,N} & \cdots  & -x_{-N,N} & p_{N}T\\
x_{1,-1} & \cdots  & x_{1,-N} & -x_{1,1} & \cdots  & -x_{N,1} & q_{1}T\\
\vdots &  & \vdots & \vdots &  & \vdots & \vdots\\
x_{1,-N} & \cdots  & x_{N,-N} & -x_{1,N} & \cdots  & -x_{N,N} & q_{N}T\\
-q_{1} & \cdots  & -q_{N} & p_{1} & \cdots  & p_{N} & 0
\end{array}
\right)  
\end{gather*}
and $C_{2N+1,2N+1}$ is the minor of $C$ obtained deleting the last
row and column. Moreover $\deg C_{2k+1}=2k+1$.
\end{proposition}

\section[The kinematical algebras in $\left(  3+1\right)  $-dimensions]{The kinematical algebras in
$\boldsymbol{\left(  3+1\right) }$-dimensions}

As an interesting physical example, we apply the preceding
procedure to the class of kinematical Lie algebras in
$(3+1)$-dimensions. This choice is appropriate since it contains
algebras of both types and because they are all related by
contractions~\cite{Ba,Lo}, thus by transitivity the procedure can
be applied. Although their invariants have been obtained
repeatedly in dif\/ferent contexts~\cite{Gro,Hp}, it is worthy to
be done on the basis of the above arguments, which, moreover, show
that the matrix providing the invariants is not necessarily
related to a faithful representation of the algebra.

Following the original notation of \cite{Ba}, kinematical Lie
algebras are def\/ined over the basis $\left\{
J_{\alpha},P_{\alpha},K_{\alpha},H\right\} _{1\leq \alpha \leq3}$,
where $J_{\alpha}$ are spatial rotations, $P_{\alpha}$ spatial
translations, $K_{\alpha}$ the boosts and $H$ the time
translation, constrained
to the condition of space isotropy%
\begin{gather*}
\left[  J_{\alpha},J_{\beta}\right]  =\varepsilon^{\alpha\beta\gamma}%
J_{\gamma}, \qquad \left[  J_{\alpha},P_{\beta}\right]
=\varepsilon^{\alpha \beta\gamma}P_{\gamma}, \qquad \left[
J_{\alpha},K_{\beta}\right]  =\varepsilon
^{\alpha\beta\gamma}K_{\gamma},\qquad \left[  J_{\alpha},H\right]
=0,
\end{gather*}
as well as the assumption that time-reversal and parity are
automorphisms of the group. Taking the compact notation $\left[
\mathbf{X},\mathbf{Y}\right] =\mathbf{Z}$ for $\left[
X_{\alpha},Y_{\beta}\right] =\varepsilon
^{\alpha\beta\gamma}Z_{\gamma}$, the brackets of the nonisomorphic
kinematical Lie algebras are given in Table 1\footnote{We have
omitted the Para--Poincar\'{e} and Para--Galilei Lie algebras,
since they are isomorphic to the Poincar\'{e} and Galilei
algebras, respectively, although they are physically dif\/ferent.
For purposes of invariants, this physical distinction is
irrelevant.}.

\begin{table}[t]
\small
\caption{Nonisomorphic kinematical algebras in $(3+1)$
dimensions~\cite{Ba}.} \vspace{2mm}

\centerline{\begin{tabular}
[c]{l|ccccccccc}%
& $\frak{so}\left( 4,1\right)  $ & $\frak{so}\left(  3,2\right) $
& $I\frak{so}\left(  3,1\right)  $ & $I\frak{so}\left(  4\right) $
& Ne$^{\exp }$ & Ne$^{\rm osc}$ & Carroll & G$\left(  2\right)  $
& Static\\\hline 
&&&&&&&&\\[-2mm]
$\left[  H,\mathbf{P}\right]  $ & $\mathbf{K}$ &
$-\mathbf{K}$ & $0$ &
$\mathbf{K}$ & $\mathbf{K}$ & $\mathbf{-K}$ & $0$ & $0$ & $0$\\
$\left[  H,\mathbf{K}\right]  $ & $\mathbf{P}$ & $\mathbf{P}$ &
$\mathbf{P}$ &
$0$ & $\mathbf{P}$ & $\mathbf{P}$ & $0$ & $\mathbf{P}$ & $0$\\
$\left[  \mathbf{P,P}\right]  $ & $\mathbf{J}$ & $\mathbf{-J}$ &
$0$ &
$\mathbf{J}$ & $0$ & $0$ & $0$ & $0$ & $0$\\
$\left[  \mathbf{K,K}\right]  $ & $\mathbf{-J}$ & $\mathbf{-J}$ &
$\mathbf{-J}$ & $0$ & $0$ & $0$ & $0$ & $0$ & $0$\\
$\left[  \mathbf{P,K}\right]  $ & $H$ & $H$ & $H$ & $H$ & $0$ &
$0$ & $H$ &
$0$ & $0$%
\end{tabular}}
\end{table}

As shown in \cite{Ba}, any kinematical Lie algebra is obtained by
contraction of the simple de~Sitter Lie algebras $\frak{so}\left(
3,2\right)  $ and $\frak{so}\left(  4,1\right)  $. With the
exception of the static algebra, all of the remaining ones possess
two independent invariants.

\subsection{De Sitter algebras}

Since both de Sitter algebras are simple and pseudo-orthogonal,
their Casimir operators follow at once from application of the
Gel'fand formula. For obtaining the invariants over the
kinematical basis above, the matrix (\ref{ST}) has to be slightly
transformed.

\begin{enumerate}\vspace{-2mm}\itemsep=0pt
\item  Anti de Sitter algebra $\frak{so}\left(  3,2\right). $ The
matrix related to the standard representations is:
\begin{gather}
D=\left(
\begin{array}
[c]{ccccc}%
0 & j_{3} & j_{2} & -k_{1} & p_{1}\\
-j_{3} & 0 & j_{1} & k_{2} & -p_{2}\\
-j_{2} & -j_{1} & 0 & -k_{3} & p_{3}\\
-k_{1} & k_{2} & -k_{3} & 0 & h\\
p_{1} & -p_{2} & p_{3} & -h & 0
\end{array}
\right). \label{AS}
\end{gather}
Computing the characteristic polynomial we have $\left| D-T\,{\rm
Id}_{5}\right| =T^{5}+C_{2}T^{3}+C_{4}T$, where
\begin{gather*}
C_{2}   =j_{\alpha}j^{\alpha}-p_{\alpha}p^{\alpha}-k_{\alpha}k^{\alpha}%
+h^{2},\\
C_{4}   =j_{\alpha}j^{\alpha}h^{2}+\left(
(p_{\alpha}p^{\alpha})(k_{\alpha}k^{\alpha})-(p_{\alpha}k^{\alpha})^2
\right)-\left( j_{\alpha}p^{\alpha}\right)  ^{2}-\left(
j_{\alpha}k^{\alpha}\right)
^{2}-2\varepsilon^{\alpha\beta\gamma}j_{\alpha
}p_{\beta}k_{\gamma}h.
\end{gather*}

\item  De Sitter algebra $\frak{so}\left(  4,1\right) $. The
resulting matrix is similar to the previous one:
\begin{gather*}
D=\left(
\begin{array}
[c]{ccccc}%
0 & j_{3} & j_{2} & -k_{1} & p_{1}\\
-j_{3} & 0 & j_{1} & k_{2} & -p_{2}\\
-j_{2} & -j_{1} & 0 & -k_{3} & p_{3}\\
-k_{1} & k_{2} & -k_{3} & 0 & h\\
-p_{1} & p_{2} & -p_{3} & h & 0
\end{array}
\right).
\end{gather*}
Then we have $\left|  D-T\,{\rm Id}_{5}\right|
=T^{5}+C_{2}T^{3}+C_{4}T$, where
\begin{gather*}
C_{2}   =j_{\alpha}j^{\alpha}+p_{\alpha}p^{\alpha}-k_{\alpha}k^{\alpha}%
-h^{2},\\
C_{4}   =-j_{\alpha}j^{\alpha}h^{2}-\left(
(p_{\alpha}p^{\alpha})(k_{\alpha}k^{\alpha})-(p_{\alpha}k^{\alpha})^2
\right)+\left( j_{\alpha}p^{\alpha}\right)  ^{2}-\left(
j_{\alpha}k^{\alpha}\right)
^{2}+2\varepsilon^{\alpha\beta\gamma}j_{\alpha
}p_{\beta}k_{\gamma}h.
\end{gather*}
\end{enumerate}

For these two algebras, the result follows at once from the
Gel'fand formula.

\subsection{The nonrelativistic cosmological Lie algebras}

The Newton algebras Ne$^{+}$ and Ne$^{-}$ are obtained as
contractions of the de Sitter and Anti de Sitter algebras,
respectively. Since the subalgebra generated by
$\left\{K_{\alpha},P_{\alpha}\right\}$ is Abelian, it will follow
that the invariants of these algebras will not depend on the
rotation and time-translation generators.

\begin{enumerate}\vspace{-2mm}\itemsep=0pt
\item  The Newton algebra Ne$^{-}$ [oscillating universe]. It can
be easily verif\/ied that this Lie \linebreak algebra is indeed an
extension by a derivation of the nine dimensional Lie
algebra\linebreak $\frak{so}\left(  3\right)
\overrightarrow{\oplus}_{2ad\frak{so}\left(  3\right)  }6L_{1}$.
By the comment above, the matrix giving the Casimir operators of
Ne$^{-}$ will not be related to a faithful representation of the
algebra. We have:
\begin{gather*}
D=\left(
\begin{array}
[c]{ccccc}%
0 & 0 & 0 & -k_{1} & p_{1}\\
0 & 0 & 0 & k_{2} & -p_{2}\\
0 & 0 & 0 & -k_{3} & p_{3}\\
-k_{1} & k_{2} & -k_{3} & 0 & 0\\
p_{1} & -p_{2} & p_{3} & 0 & 0
\end{array}
\right).
\end{gather*}
Expanding the secular equation we arrive at $\left| D-T\,{\rm
Id}_{5}\right| =T^{5}+C_{2}T^{3}+C_{4}T$, where
\begin{gather*}
C_{2}   =-p_{\alpha}p^{\alpha}-k_{\alpha}k^{\alpha},\\ C_{4}
=\left(
(p_{\alpha}p^{\alpha})(k_{\alpha}k^{\alpha})-(p_{\alpha}k^{\alpha})^2
\right).
\end{gather*}

\item  The Newton algebra Ne$^{+}$ [expanding universe]. This Lie
algebra is also an extension by a~derivation of $\frak{so}\left(
3\right) \overrightarrow{\oplus}_{2ad\frak{so}\left(  3\right)
}6L_{1}$. In this case the matrix to be used is:
\begin{gather*}
D=\left(
\begin{array}
[c]{ccccc}%
0 & 0 & 0 & -k_{1} & p_{1}\\
0 & 0 & 0 & k_{2} & -p_{2}\\
0 & 0 & 0 & -k_{3} & p_{3}\\
-k_{1} & k_{2} & -k_{3} & 0 & 0\\
-p_{1} & p_{2} & -p_{3} & 0 & 0
\end{array}
\right).
\end{gather*}
Then we have $\left|  D-T\,{\rm Id}_{5}\right|
=T^{5}+C_{2}T^{3}+C_{4}T$, where
\begin{gather*}
C_{2}   =p_{\alpha}p^{\alpha}-k_{\alpha}k^{\alpha},\\
C_{4}
=\left(
(p_{\alpha}p^{\alpha})(k_{\alpha}k^{\alpha})-(p_{\alpha}k^{\alpha})^2
\right).
\end{gather*}
\end{enumerate}

\subsection{The inhomogeneous (pseudo)-orthogonal algebras}

In order to obtain the matrix for the inhomogeneous algebras $I\frak{so}%
\left(  3,1\right)  $ and $I\frak{so}\left(  4\right)  $, we use
the contractions $\frak{so}\left(  3,2\right)  \rightsquigarrow
I\frak{so}\left( 3,1\right)  $ and $\frak{so}\left(  4,1\right)
\rightsquigarrow I\frak{so}\left(  4\right)  $, respectively.

\begin{enumerate}\vspace{-2mm}\itemsep=0pt
\item  The Poincar\'{e} Lie algebra $I\frak{so}\left(  3,1\right)
$. The matrix $D$ is given by
\begin{gather*}
D=\left(
\begin{array}
[c]{ccccc}%
0 & j_{3} & j_{2} & -k_{1} & p_{1}\\
-j_{3} & 0 & j_{1} & k_{2} & -p_{2}\\
-j_{2} & -j_{1} & 0 & -k_{3} & p_{3}\\
-k_{1} & k_{2} & -k_{3} & 0 & h\\
p_{1} & -p_{2} & p_{3} & -h & 0
\end{array}
\right).
\end{gather*}
We  def\/ine $P\left(  T\right)  =\left|  D-T\,{\rm Id}_{5}\right|
+T$ $\left| D_{55}-T\,{\rm Id}_{4}\right|  $, where $D_{55}$ is
the minor of $D$ obtained by deleting the f\/ifth column and row.
In particular, the matrix $D_{55}$ corresponds to that of the
Lorentz algebra $\frak{so}\left(  3,1\right)  $. Expanding the
expression for $P\left( T\right)  $, we get  $P\left(  T\right)
=T^{5}+C_{2}T^{3}+C_{4}T$, where
\begin{gather*}
C_{2}   =h^{2}-p_{\alpha}p^{\alpha},\\
C_{4}   =j_{\alpha}j^{\alpha}h^{2}+\left(
p_{\alpha}p^{\alpha}\right) \left(  k_{\alpha}k^{\alpha}\right)
-\left(  j_{\alpha}p^{\alpha}\right) ^{2}-\left(
p_{\alpha}k^{\alpha}\right)  ^{2}-2\varepsilon^{\alpha\beta
\gamma}j_{\alpha}p_{\beta}k_{\gamma}h.
\end{gather*}
Moreover, the matrix $D$ can be decomposed as
\[
D=D_{1}+D_{2}=\left(
\begin{array}
[c]{ccccc}%
0 & j_{3} & j_{2} & -k_{1} & p_{1}\\
-j_{3} & 0 & j_{1} & k_{2} & -p_{2}\\
-j_{2} & -j_{1} & 0 & -k_{3} & p_{3}\\
-k_{1} & k_{2} & -k_{3} & 0 & h\\
0 & 0 & 0 & 0 & 0
\end{array}
\right)  +\left(
\begin{array}
[c]{ccccc}%
0 & 0 & 0 & 0 & 0\\
0 & 0 & 0 & 0 & 0\\
0 & 0 & 0 & 0 & 0\\
0 & 0 & 0 & 0 & 0\\
p_{1} & -p_{2} & p_{3} & -h & 0
\end{array}
\right)  ,
\]
where $D_{1}$ def\/ines a faithful representation of
$I\frak{so}\left( 3,1\right)  $.

\item  The inhomogeneous algebra $I\frak{so}\left(  4\right) $.
Here the polynomial is $P\left(  T\right)  =\left|  D-T\,{\rm
Id}_{5}\right|  +T|  D_{55} -{}$ $T\,{\rm Id}_{4}|  $, where
$D_{55}$ is the minor of $D$ obtained deleting the f\/ifth column
and row. The matrix~$D$ is given by
\begin{gather*}
D=\left(
\begin{array}
[c]{ccccc}%
0 & j_{3} & j_{2} & -k_{1} & p_{1}\\
-j_{3} & 0 & j_{1} & k_{2} & -p_{2}\\
-j_{2} & -j_{1} & 0 & -k_{3} & p_{3}\\
-k_{1} & k_{2} & -k_{3} & 0 & h\\
-p_{1} & p_{2} & -p_{3} & h & 0
\end{array}
\right).
\end{gather*}
Expanding $P\left( T\right)  $, we get $P\left( T\right)
=T^{5}+C_{2}T^{3}+C_{4}T$, with
\begin{gather*}
C_{2}   =-h^{2}-k_{\alpha}k^{\alpha},\\
C_{4}   =j_{\alpha}j^{\alpha}h^{2}+\left(
p_{\alpha}p^{\alpha}\right) \left(  k_{\alpha}k^{\alpha}\right)
+\left(  j_{\alpha}k^{\alpha}\right) ^{2}-\left(
p_{\alpha}k^{\alpha}\right)  ^{2}-2\varepsilon^{\alpha\beta
\gamma}j_{\alpha}p_{\beta}k_{\gamma}h.
\end{gather*}
\end{enumerate}
In this case, $D$ decomposes as
\[
D=D_{1}+D_{2}=\left(
\begin{array}
[c]{ccccc}%
0 & j_{3} & j_{2} & 0 & p_{1}\\
-j_{3} & 0 & j_{1} & 0 & -p_{2}\\
-j_{2} & -j_{1} & 0 & 0 & p_{3}\\
-k_{1} & k_{2} & -k_{3} & 0 & h\\
-p_{1} & p_{2} & -p_{3} & 0 & 0
\end{array}
\right)  +\left(
\begin{array}
[c]{ccccc}%
0 & 0 & 0 & -k_{1} & 0\\
0 & 0 & 0 & k_{2} & 0\\
0 & 0 & 0 & -k_{3} & 0\\
0 & 0 & 0 & 0 & 0\\
0 & 0 & 0 & h & 0
\end{array}
\right)  ,
\]
and $D_{1}$ is the matrix related to a the faithful representation
of $I\frak{so}\left(  4\right)  $ by $5\times5$ matrices.

\subsection{The Carroll Lie algebra}

Among the classical kinematical Lie algebras, the Carroll Lie
algebra is the only isomorphic to the semidirect product of a
simple Lie algebra (the compact algebra $\frak{so}\left(  3\right)
$) and a Heisenberg Lie algebra. Indeed the noncentral Casimir
operator can be determined using the determinant procedure
developed in \cite{C42}. However, the Casimir operators (the
second is trivial, since the centre is nonzero) can also be
obtained by the same method as before.

We def\/ine $P\left(  T\right)  =\left| D-T\,{\rm Id}_{5}\right|$,
where
 $D$ is the matrix
\begin{gather*}
D=\left(
\begin{array}
[c]{ccccc}%
0 & j_{3} & j_{2} & -k_{1} & p_{1}\\
-j_{3} & 0 & j_{1} & k_{2} & -p_{2}\\
-j_{2} & -j_{1} & 0 & -k_{3} & p_{3}\\
-k_{1} & k_{2} & -k_{3} & T & h\\
-p_{1} & p_{2} & -p_{3} & h & T
\end{array}
\right).
\end{gather*}
Observe that in this case, the matrix $D$ is dependent on the
variable $T$. Expanding, we get  $P\left(  T\right)
=T^{5}+C_{2}T^{3}+C_{4}T$, where
\begin{gather*}
C_{2}   =h^{2},\\ C_{4}   =j_{\alpha}j^{\alpha}h^{2}+\left(
p_{\alpha}p^{\alpha}\right) \left(  k_{\alpha}k^{\alpha}\right)
-\left(  p_{\alpha}k^{\alpha}\right)
^{2}-2\varepsilon^{\alpha\beta\gamma}j_{\alpha}p_{\beta}k_{\gamma}h.
\end{gather*}
The matrix $D$ decomposes in this case as
\[
D=D_{1}+D_{2}=\left(
\begin{array}
[c]{ccccc}%
0 & j_{3} & j_{2} & 0 & p_{1}\\
-j_{3} & 0 & j_{1} & 0 & -p_{2}\\
-j_{2} & -j_{1} & 0 & 0 & p_{3}\\
-k_{1} & k_{2} & -k_{3} & 0 & h\\
0 & 0 & 0 & 0 & 0
\end{array}
\right)  +\left(
\begin{array}
[c]{ccccc}%
0 & 0 & 0 & -k_{1} & 0\\
0 & 0 & 0 & k_{2} & 0\\
0 & 0 & 0 & -k_{3} & 0\\
0 & 0 & 0 & T & 0\\
-p_{1} & p_{2} & -p_{3} & h & T
\end{array}
\right).
\]
Again, the matrix $D_{1}$ gives rise to a faithful representation
of the Carroll algebra.

\subsection[The Galilei algebra $G(2)$]{The Galilei algebra $\boldsymbol{G(2)}$}

As happened for the Newton algebras, the Casimir operators of the
Galilei algebra do not depend on the variables
$\left\{j_{\alpha},h\right\}$. Here we consider the polynomial
$P\left(  T\right)  =\left|  D-T\,{\rm Id}_{5}\right|$, where
\begin{gather*}
D=\left(
\begin{array}
[c]{ccccc}%
0 & 0 & 0 & -k_{1} & p_{1}\\
0 & 0 & 0 & k_{2} & -p_{2}\\
0 & 0 & 0 & -k_{3} & p_{3}\\
-k_{1} & k_{2} & -k_{3} & 0 & 0\\
-p_{1} & p_{2} & -p_{3} & 0 & T
\end{array}
\right).
\end{gather*}
Also in this case, the matrix $D$ is dependent on the variable
$T$. Developing the polynomial we obtain  $P\left(  T\right)
=T^{5}+C_{2}T^{3}+C_{4}T$, where
\begin{gather*}
C_{2}   =p_{\alpha}p^{\alpha},\\
C_{4}   =\left(
p_{\alpha}p^{\alpha}\right)  \left( k_{\alpha}k^{\alpha }\right)
-\left(  p_{\alpha}k^{\alpha}\right)  ^{2}.
\end{gather*}
By the remark above, the preceding matrix is not related to a
faithful representation of the algebra.

\subsection{The static Lie algebra}

This algebra is nothing but the splittable af\/f\/ine Lie algebra
$\big( \frak{so}\left(  3\right)
\overrightarrow{\oplus}_{2ad\frak{so}\left( 3\right)
}6L_{1}\big)  \oplus\mathbb{R}$. As commented, it has four
invariants, all of the degree two,
\[
I_{1}=h,\qquad I_{2}=p_{\alpha}p^{\alpha},\qquad I_{3}=k_{\alpha}k^{\alpha}%
,\qquad I_{5}=k_{\alpha}p^{\alpha}.
\]
To obtain them in matrix form, we consider
\begin{gather*}
D=\left(
\begin{array}
[c]{ccccc}%
0 & 0 & 0 & -k_{1} & p_{1}T\\
0 & 0 & 0 & k_{2} & -p_{2}T\\
0 & 0 & 0 & -k_{3} & p_{3}T\\
-k_{1} & k_{2} & -k_{3} & 0 & -hT\\
-p_{1} & p_{2} & -p_{3} & -h & 0
\end{array}
\right)
\end{gather*}
and obtain  $P\left(  T\right)  =T^{5}+\left(
I_{2}-I_{1}^{2}\right) T^{4}-I_{3}T^{3}-\left(
I_{2}I_{3}-I_{5}^{2}\right)  T^{2}$. Simplifyng the
coef\/f\/icients we recover the basis of invariants above. Since
the variables associated to the rotations $J_{\alpha}$ do not
appear in the invariants, the matrix $D$ does not provide a
representation of the static algebra.

For later use we consider the following functions:
\begin{gather}
I_{1}=h,\qquad I_{2}=p_{\alpha}p^{\alpha},\qquad I_{3}=k_{\alpha}k^{\alpha}%
,\qquad I_{4}=j_{\alpha}j^{\alpha},\qquad I_{5}=k_{\alpha}p^{\alpha},\nonumber\\
 I_{6}=j_{\alpha
}k^{\alpha},\qquad I_{7}=j_{\alpha}p^{\alpha},\qquad
M=\varepsilon^{\alpha\beta\gamma}j_{\alpha}p_{\beta}k_{\gamma}.
\label{FS}
\end{gather}

\section{Applications: The missing label problem}

As known, irreducible representations of a semisimple Lie algebra
are labelled unambigously by the eigenvalues of Casimir operators.
In a more general frame, irreducible representations of a Lie
algebra $\frak{g}$ are labelled usingby means of the eigenvalues
of its generalized Casimir invariants~\cite{Sh2,Sh}. The number of
internal labels needed is
\begin{gather*}
i=\frac{1}{2}(\dim \frak{g}- \mathcal{N}(\frak{g})).
\end{gather*}
If we use a subalgebra $\frak{h}$ label the basis states of
$\frak{g}$, then we obtain $\frac{1}{2}(\dim
\frak{h}+\mathcal{N}(\frak{h}))+l^{\prime}$ labels from~$\frak{h}$,
where $l^{\prime}$ is the number of invariants of $\frak{g}$ that
depend only on variables of the subalgebra~$\frak{h}$~\cite{Sh}.
In order to label irreducible representations of $\frak{g}$
uniquely, it is therefore necessary to f\/ind
\begin{gather}
n=\frac{1}{2}\left( \dim\frak{g}-\mathcal{N}(\frak{g})
-\dim\frak{h}-\mathcal{N}(\frak{h})\right)+l^{\prime} \label{ML}
\end{gather}
additional operators, which are usually called missing label
operators. They are found by integra\-ting the equations of system
(\ref{sys}) corresponding to the subalgebra generators. The total
number of available operators of this kind is easily shown to be
$m=2n$.

In this situation, it is plausible to think that whenever the
Casimir operators of a Lie algebra~$\frak{g}$ can be determined
using determinants of (polynomial) matrices, the same procedure
could hold for computing missing label operators according to some
subalgebra $\frak{h}$. In this section we analyze the missing
label problem for the chain
\[
\frak{so}(3)\hookrightarrow \frak{g},
\]
where $\frak{g}$ is a kinematical Lie algebra in
$(3+1)$-dimensions and  $\frak{so}\left(  3\right)  $ the compact
subalgebra generated by the $\left\{  J_{\mu\nu}\right\}  $. The
missing label operators are among the solutions of the equations:
\begin{gather*}
j_{3}\frac{\partial F}{\partial j_{2}}-j_{2}\frac{\partial F}{\partial j_{3}%
}+p_{3}\frac{\partial F}{\partial p_{2}}-p_{2}\frac{\partial
F}{\partial p_{3}}+k_{3}\frac{\partial F}{\partial
k_{2}}-k_{2}\frac{\partial F}{\partial k_{3}}   =0,\\
-j_{3}\frac{\partial F}{\partial j_{1}}+j_{1}\frac{\partial F}{\partial j_{3}%
}-p_{3}\frac{\partial F}{\partial p_{1}}+p_{1}\frac{\partial
F}{\partial p_{3}}-k_{3}\frac{\partial F}{\partial
k_{1}}+k_{1}\frac{\partial F}{\partial
k_{3}}   =0,\\
j_{2}\frac{\partial F}{\partial j_{1}}-j_{1}\frac{\partial F}{\partial j_{2}%
}+p_{2}\frac{\partial F}{\partial p_{1}}-p_{1}\frac{\partial
F}{\partial p_{2}}+k_{2}\frac{\partial F}{\partial
k_{1}}-k_{1}\frac{\partial F}{\partial k_{2}}   =0.
\end{gather*}
Due to the space isotropy condition, the above system is valid for
any kinematical Lie algebra. According to formula (\ref{ML}),
there are
\[
n=\frac{1}{2}\left(  \dim \frak{g}-\mathcal{N}\left( \frak{g}
\right) -\dim\frak{so}\left( 3\right) -\mathcal{N}\left(  so\left(
3\right)  \right)  \right)
+l^{\prime}=\frac{1}{2}\left(6-\mathcal{N}(\frak{g})\right)+l^{\prime}
\]
missing labels. In any case we have $l^{\prime}=0$. Moreover, for
any kinematical algebra, with the exception of the static algebra,
we obtain $n=2$ and four available missing label operators, while
for the static algebra we get $n=1$ and $m=2$. By using of the
matrix notation, the system can be rewritten as:
\begin{gather}
\left(
\begin{array}
[c]{cccccccccc}%
0 & j_{3} & -j_{2} & 0 & p_{3} & -p_{2} & 0 & k_{3} & -k_{2} & 0\\
-j_{3} & 0 & j_{1} & -p_{3} & 0 & p_{1} & -k_{3} & 0 & k_{1} & 0\\
j_{2} & -j_{1} & 0 & p_{2} & -p_{1} & 0 & k_{2} & -k_{1} & 0 & 0
\end{array}
\right)  \left(
\begin{array}[c]{c}%
\frac{\partial F}{\partial j_{\alpha}}\vspace{2mm}\\
\frac{\partial F}{\partial p_{\alpha}}\vspace{2mm}\\
\frac{\partial F}{\partial k_{\alpha}}\vspace{2mm}\\
\frac{\partial F}{\partial h}
\end{array}
\right)  =0.\label{S1}
\end{gather}
Since the matrix has rank three, there are seven independent
solutions of the system. The number of solutions that do not
depend on the variables $\left\{ j_{\alpha}\right\}  $ of the
subalgebra $\frak{so}\left(  3\right)  $ is given by \cite{C35}:
\[
\mathcal{N}^{\prime}=7-{\rm rank}\left(
\begin{array}
[c]{ccccccc}%
0 & p_{3} & -p_{2} & 0 & k_{3} & -k_{2} & 0\\
-p_{3} & 0 & p_{1} & -k_{3} & 0 & k_{1} & 0\\
p_{2} & -p_{1} & 0 & k_{2} & -k_{1} & 0 & 0
\end{array}
\right)  =4.
\]
It is straightforward to verify that a complete system of
independent solutions is given by:
\begin{gather}
\left\{I_{1}=h,\ I_{2}=p_{\alpha}p^{\alpha},\ I_{3}=k_{\alpha}k^{\alpha}%
,\ I_{4}=j_{\alpha}j^{\alpha},\ I_{5}=k_{\alpha}p^{\alpha},\
I_{6}=j_{\alpha }k^{\alpha},\ I_{7}=j_{\alpha}p^{\alpha}\right\}.
\label{FS1}
\end{gather}
In particular, the invariants $I_{1}$, $I_{2}$, $I_{3}$, $I_{5}$,
which are the independent solutions not involving the variables
$j_{\alpha}$, constitute a set of solutions for the static Lie
algebra. The function
$M=\varepsilon^{\alpha\beta\gamma}j_{\alpha}p_{\beta}k_{\gamma}$
of~(\ref{FS}) is functionally dependent on the previous functions,
as shown by the relation
\begin{gather*}
M^{2}=I_{5}^{2}I_{4}+I_{7}^{2}I_{3}-I_{6}^{2}I_{2}-I_{2}I_{3}I_{4}-2I_{5}I_{6}I_{7}.
\end{gather*}
To see how the matrices used for the Casimir operators of
kinematical algebras can also be applied to the MLP, we consider
the Anti de Sitter algebra $\frak{so}(3,2)$. In the notation of
(\ref{FS1}), the invariants of the algebra are given by
$C_{2}=I_{1}^{2}-I_{2}-I_{3}+I_{4}$ and
$C_{4}=I_{1}^{2}I_{4}+I_{2}I_{3}-I_{5}^{2}-I_{6}^{2}-I_{7}^{2}-2I_{1}M$,
while $I_{4}$ clearly represents the Casimir operator of
$\frak{so}(3)$. We now look for those missing label operators that
depend only on the variables
$\left\{p_{\alpha},k_{\alpha},h\right\}$. To this extent, we
consider the matrix (\ref{AS}) used to compute $C_{2}$ and $C_{4}$
and replace the variables $j_{\alpha}$ by $0$. We obtain
\begin{gather*}
D^{\prime}=\left(
\begin{array}
[c]{ccccc}%
0 & 0 & 0 & -k_{1} & p_{1}\\
0 & 0 & 0 & k_{2} & -p_{2}\\
0 & 0 & 0 & -k_{3} & p_{3}\\
-k_{1} & k_{2} & -k_{3} & 0 & h\\
p_{1} & -p_{2} & p_{3} & -h & 0
\end{array}
\right).
\end{gather*}
Considering the characteristic polynomial we have
\[
P\left(T\right)=\left|D^{\prime}-T\,{\rm
Id}_{5}\right|=T^{5}+(I_{1}^{2}-I_{2}-I_{3})T^{3}+(I_{2}I_{3}-I_{5}^{2})T.
\]
It can be easily verif\/ied that
$\left\{I_{2},I_{3},I_{5}\right\}$ are solutions of (\ref{S1})
independent of $\left\{C_{2},C_{4},I_{4}\right\}$, while
$I_{1}^{2}$ is not an independent solution. Therefore the matrix
$D^{\prime}$ provides three of the four available missing label
operators. The fourth, which can be chosen as~$I_{6}$, cannot be
obtained using~$D^{\prime}$, since it depends on the rotation
generators.

A similar argument can be used for the remaining kinematical
algebras $\frak{g}$. We consider the matrix giving the invariants
of $\frak{g}$ and replace the subalgebra generators $j_{\alpha}$
by $0$. Then we compute the corresponding characteristic
polynomial and see how many independent solutions from
$\left\{C_{2},C_{4},I_{4}\right\}$ are obtained, where $C_{2}$ and
$C_{4}$ are the quadratic and fourth order Casimir operators of
$\frak{g}$, respectively. Only for the static Lie algebra this
method of generating missing label operators fails, since their
Casimir operators are a maximal set of solutions of system
(\ref{S1}) not depending on the $j_{\alpha}$. The corresponding
results are presented in Table~2.

\begin{table}[t]\small
\caption{Missing label operators for the chain
$\frak{so}(3)\hookrightarrow \frak{g}$.} \vspace{2mm}

\centerline{
\begin{tabular}
[c]{l|lll}%
Algebra $\frak{g}$ & Casimir operators of $\frak{g}$ &
\begin{tabular}
[c]{l}%
Missing label \\
operators
\end{tabular}
&
\begin{tabular}
[c]{l}%
MLP obtained from\\
the reduced matrix
\end{tabular}
\\\hline
&&&\\[-3mm]
$\frak{so}\left( 4,1\right)  $ &
\begin{tabular}
[c]{l}%
$I_{1}^{2}-I_{2}+I_{3}-I_{4}$\\
$I_{1}^{2}I_{4}+I_{2}I_{3}-I_{5}^{2}+I_{6}^{2}-I_{7}^{2}-2I_{1}M$%
\end{tabular}
& $\left\{  I_{2},I_{3},I_{5},I_{6}\right\}  $ & $\left\{  I_{2},I_{3}%
,I_{5}\right\}  $\\[3mm]
$\frak{so}\left(  3,2\right)  $ &
\begin{tabular}
[c]{l}%
$I_{1}^{2}-I_{2}-I_{3}+I_{4}$\\
$I_{1}^{2}I_{4}+I_{2}I_{3}-I_{5}^{2}-I_{6}^{2}-I_{7}^{2}-2I_{1}M$%
\end{tabular}
& $\left\{  I_{2},I_{3},I_{5},I_{6}\right\}  $ & $\left\{  I_{2},I_{3}%
,I_{5}\right\}  $\\[3mm]
$I\frak{so}\left(  3,1\right)  $ &
\begin{tabular}
[c]{l}%
$I_{1}^{2}-I_{2}$\\
$I_{1}^{2}I_{4}+I_{2}I_{3}-I_{5}^{2}-I_{7}^{2}-2I_{1}M$%
\end{tabular}
& $\left\{  I_{2},I_{3},I_{5},I_{7}\right\}  $ & $\left\{  I_{2},I_{3}%
,I_{5}\right\}  $\\[3mm]
$I\frak{so}\left(  4\right)  $ &
\begin{tabular}
[c]{l}%
$I_{1}^{2}+I_{2}$\\
$I_{1}^{2}I_{4}+I_{2}I_{3}-I_{5}^{2}+I_{6}^{2}-2I_{1}M$%
\end{tabular}
& $\left\{  I_{2},I_{3},I_{5},I_{6}\right\}  $ & $\left\{  I_{2},I_{3}%
,I_{5}\right\}  $\\[3mm]
Ne$^{+}$ &
\begin{tabular}
[c]{l}%
$I_{2}-I_{3}$\\
$I_{2}I_{3}-I_{5}^{2}$%
\end{tabular}
& $\left\{  I_{1},I_{2},I_{6},I_{7}\right\}  $ & $\left\{  I_{2}\right\}  $\\
Ne$^{-}$ &
\begin{tabular}
[c]{l}%
$I_{2}+I_{3}$\\
$I_{2}I_{3}-I_{5}^{2}$%
\end{tabular}
& $\left\{  I_{1},I_{2},I_{6},I_{7}\right\}  $ & $\left\{  I_{2}\right\}  $\\
Carroll &
\begin{tabular}
[c]{l}%
$I_{1}^{2}$\\
$I_{1}^{2}I_{4}+I_{2}I_{3}-I_{5}^{2}-2I_{1}M$%
\end{tabular}
& $\left\{  I_{2},I_{3},I_{5},I_{6}\right\}  $ & $\left\{
I_{2}I_{3}\right\}
$\\
Galilei &
\begin{tabular}
[c]{l}%
$I_{2}$\\
$I_{2}I_{3}-I_{5}^{2}$%
\end{tabular}
& $\left\{  I_{1},I_{3},I_{6},I_{7}\right\}  $ & $\left\{  I_{2}I_{3}%
-I_{5}^{2}\right\}  $\\
Static & $I_{1},I_{2},I_{3},I_{5}$ & $\left\{  I_{6},I_{7}\right\}
$ & ---
\end{tabular}}
\end{table}

\section{Final remarks}

The approach presented here to compute Casimir invariants of Lie
algebras tries to extend the classical results established for
semisimple algebras to their contractions, making use of the
standard representation of the contracted algebra. Although it
provides in many cases closed formulae of the invariants of
contractions, the application of the Gel'fand formula is certainly
only of interest when the contraction has the same number of
invariants than the original algebra. Although in the algebras
analyzed here no dependence problems have been encountered, one of
the unsolved problems is to f\/ind suf\/f\/iciency criteria to
ensure that the contraction of independent invariants of an
algebra provides also independent operators of the contraction.
Work in this direction is in progress.

As concerns applications, we have seen that the MLP can be
analyzed via the generalization of the Gel'fand method, by simply
reducing the matrices by zeros corresponding to generators of the
subalgebra considered. This point of view could also be
interesting in combination with problems in symmetries of
dif\/ferential equations related to contractions of Lie algebras,
such as the separation of variables~\cite{Wi2}.

\subsection*{Acknowledgements}
The author wishes to express his gratitude to J. L\^{o}hmus for
drawing his attention to reference~\cite{Gro} and useful comments,
and to the referee for multiple suggestions that helped to improve
the manuscript. This work was supported by the research grant
PR1/05-13283 of the UCM.

\LastPageEnding

\end{document}